\documentclass[10pt,conference]{IEEEtran}
\IEEEoverridecommandlockouts

\usepackage{cite}
\usepackage[hyphens]{url}
\usepackage{tikz}
\usepackage{braket}
\usepackage{xcolor}
\usepackage{amsmath}
\usepackage{balance}
\usepackage{environ}
\usepackage{physics}
\usepackage{colortbl}
\usepackage{enumitem}
\usepackage{amsfonts}
\usepackage{booktabs}
\usepackage{graphicx}
\usepackage{textcomp}
\usepackage{algorithm}
\usepackage{algpseudocode}
\usepackage[normalem]{ulem}

\newcommand{\sol}{PaQit}

\def\BibTeX{{\rm B\kern-.05em{\sc i\kern-.025em b}\kern-.08em
    T\kern-.1667em\lower.7ex\hbox{E}\kern-.125emX}}

\begin{document}

\title{\sol{}: Energy-Runtime-Fidelity Co-Optimization for Neutral Atom Quantum Computers}

\author{
\IEEEauthorblockN{Jason Ludmir}
\IEEEauthorblockA{\textit{Rice University}\\ Houston, TX, USA\\ \texttt{jzl2@rice.edu}}
\and
\IEEEauthorblockN{Tirthak Patel}
\IEEEauthorblockA{\textit{Rice University}\\ Houston, TX, USA\\ \texttt{tp53@rice.edu}}
}

\maketitle

\begin{abstract}

Neutral-atom quantum computers provide a scalable platform for large-scale quantum computation due to their all-optical control, room-temperature operation, and flexible lattice geometry. Although the favorable energy characteristics of these systems are well recognized, the relationship between system-level energy consumption, runtime, and computational fidelity remains poorly understood, limiting practical scheduling decisions. In this work, we develop a hardware-grounded analytical model that captures how energy and runtime scale with qubit utilization in neutral-atom systems.

We introduce \sol{}, a fidelity-aware qubit packing framework that integrates device-level Rydberg interaction physics with system-level scheduling to jointly optimize energy, runtime, and fidelity. By translating fidelity targets into packing decisions, \sol{} identifies operating regimes that maximize parallelism while respecting interaction-driven crosstalk constraints. We validate the analytical framework using simulations of QuEra's analog Aquila system and digital Gemini system, as well as real hardware executions, demonstrating close agreement between the predicted trends and observed system behavior.

\end{abstract}


\section{Introduction}
\label{sec:introduction}

Quantum computing has reached a stage where hardware scalability and energy efficiency are becoming central design concerns~\cite{auffeves2022quantum,cooper2022quantum,arora2024sustainable,jaschke2023quantum,boger2023dual,Boger_2025,Pasqal_2023}. As experimental platforms increase qubit counts and infrastructural resources, energy use directly impacts operational cost, throughput, and thermal stability~\cite{deng2024power,stade2025routing,sunami2025scalable}. However, most existing work focuses on algorithmic fidelity, compilation, or control optimization, without characterizing how energy, runtime, and fidelity jointly influence scheduling decisions. Although the favorable energy characteristics of quantum computing have been discussed~\cite{jaschke2023quantum}, there is currently no framework that translates these system-level trade-offs into practical scheduling policies. Understanding this relationship is critical for designing quantum systems that are both computationally effective and energy-efficient.

Neutral-atom (also known as Rydberg-atom) quantum computers provide a compelling platform for such system-level analysis~\cite{balewski2024engineering}. Their all-optical trapping, low-power control, reconfigurable initialization, and atom movement, combined with the absence of cryogenic infrastructure, enable inherently low-power operation even at large qubit scales~\cite{weiss2017quantum,xu2024constant}. Despite these advantages, the total system power consumption and its dependence on qubit utilization, control fields, and supporting electronics remain analytically uncharacterized. Without such models, it is difficult to reason about key system-level trade-offs between execution time, spatial utilization, and energy efficiency. We aim to address this problem.

\vspace{2mm}

\noindent\textbf{Contributions of This Work.} We present a hardware-aware systems framework, called \sol{}\footnote{This work is published at the IEEE International Conference on Quantum Computing and Engineering (QCE), 2026.}, that translates device-level Rydberg interaction physics into a practical qubit packing policy for jointly optimizing energy, runtime, and fidelity in neutral atom quantum computers.

\begin{itemize}

    \item \textbf{First-principles Energy and Runtime Modeling.} We develop a physics-grounded model that connects Rydberg atom interaction dynamics to macroscopic power behavior. Using this model, we derive analytical expressions for total runtime and energy consumption under serial and maximally parallel execution regimes, establishing fundamental upper and lower bounds.

    \vspace{1mm}

    \item \textbf{Characterization of Packing-induced Fidelity Degradation.} We show that while increasing spatial packing improves throughput and energy efficiency, it introduces interaction-driven crosstalk that degrades fidelity. This reveals an inherent trade-off between qubit utilization and computational accuracy in neutral atom systems.

    \vspace{1mm}

    \item \textbf{\sol{}: Fidelity-aware Qubit Packing for Efficiency.} We propose \sol{}, a hardware-aware qubit packing framework that leverages the distance-dependent decay of Rydberg interactions to control spatial utilization. By tuning packing density as a function of target fidelity, \sol{} identifies operating regimes that jointly optimize energy, runtime, and fidelity.

    \vspace{1mm}

    \item \textbf{Validation on Neutral Atom Systems.} We validate our analytical models and \sol{} on simulations of QuEra’s analog Aquila system~\cite{wurtz2306aquila} and digital Gemini system~\cite{Sales_Rodriguez_2025}. Our results demonstrate agreement between the predicted trends and observed system behavior, and show that fidelity-aware packing can significantly reduce total energy while maintaining stable performance. For example, we observe that spacing beyond $10\,\mu$m incurs negligible fidelity loss on Aquila, enabling efficient parallel execution. We also validate these results on the QuEra Aquila using real hardware.

    \vspace{1mm}

    \item \textbf{\sol{}'s Code and Data.} \sol{} is available at \texttt{\url{https://github.com/positivetechnologylab/PaQit}}.
    
\end{itemize}

\textit{To our knowledge, \sol{} is the first work to formally connect energy, runtime, and fidelity in Rydberg-atom quantum computers, providing a systems-level foundation for energy-efficient quantum operation.}
\section{Background and Motivation}

\noindent\textbf{Brief Fundamentals.} Neutral atom quantum computers encode qubits in the electronic states of laser-cooled atoms confined in optical lattices or programmable tweezer arrays. Each atom represents a qubit, typically using a ground state $\ket{g}$ and a Rydberg-excited state $\ket{r}$, separated by an optical transition frequency~\cite{wang2024atomique,bluvstein2022quantum,bluvstein2021controlling}. Quantum gates are implemented through optical control pulses that drive coherent transitions between these states, while entangling interactions arise from the strong dipole-dipole coupling between atoms simultaneously excited to Rydberg states. Because confinement and control are achieved entirely optically, these systems operate at or near room temperature, eliminating the need for cryogenic infrastructure and enabling scalable deployment~\cite{wurtz2306aquila}.

The dynamics of such systems are governed by the following time-dependent Hamiltonian:

\begin{equation}
\begin{split}
H(t) = \frac{\Omega(t)}{2} \sum_i \left( e^{i\phi(t)} |g\rangle_i \langle r|_i + e^{-i\phi(t)} |r\rangle_i \langle g|_i \right) \\
- \Delta(t) \sum_i \hat{n}_i + \sum_{i < j} \frac{C_6}{|\vec{p}_i - \vec{p}_j|^6} \hat{n}_i \hat{n}_j,
\end{split}
\label{eq:rydberg_ham}
\end{equation}
where $\Omega(t)$ is the global Rabi frequency, $\phi(t)$ is the phase, and $\Delta(t)$ is the detuning of the applied laser field. The operator $\hat{n}_i = |r\rangle_i\langle r|_i$ counts Rydberg excitations, and the final term captures the pairwise van der Waals interaction between atoms $i$ and $j$, parameterized by $C_6$ and their separation $d = |\vec{p}_i - \vec{p}_j|$~\cite{dibrita2025resq,dibrita2024recon}. This interaction decays as $1/r_{ij}^6$ and fundamentally governs both entanglement generation and cross-qubit interference within this system.

The dominant interaction mechanism is the Rydberg blockade effect. When two atoms lie within a blockade radius $r_b$, excitation of one atom shifts the energy levels of its neighbors, suppressing simultaneous excitation. This effect enables fast, high-fidelity entangling gates, but also introduces sensitivity to spatial configuration, making interaction-induced crosstalk a key system-level consideration.

\vspace{2mm}

\noindent\textbf{Power Consumption.} From a systems engineering perspective, total power consumption in neutral atom quantum computers arises from three primary subsystems: (1) optical trapping infrastructure that maintains atom confinement within the optical tweezers, (2) global and local laser systems that implement control and gate operations, and (3) electronic control hardware responsible for waveform generation, feedback, and measurement~\cite{bluvstein2025architectural,tan2025compilation,wang2024q}. 

The trapping infrastructure provides a largely constant baseline power $P_0$, independent of the number of active qubits, while the control and electronic subsystems scale with the number of atoms being driven and the experiment's temporal duty cycle. Importantly, recent neutral-atom platforms exhibit highly favorable scaling behavior, with incremental power per qubit remaining small relative to the static baseline. This creates a regime in which system power is dominated by static overhead, thereby fundamentally distinguishing neutral-atom architectures from other quantum computing platforms.

\vspace{2mm}

\noindent\textbf{Parallel Execution Opportunity.} Many quantum algorithms, particularly variational and sampling-based workloads, require repeated execution (``shots'') of the same quantum circuit. In neutral atom systems, these identical circuit instances can be executed concurrently within a single physical array under a shared Hamiltonian evolution, without requiring multiplexed program execution~\cite{ludmir2024modeling,ludmir2024parallax}. This property enables a form of intra-device parallelism that is not directly analogous to classical multi-programming, where different workloads compete for shared resources (in the quantum case, instances of the same program are co-run).

By spatially distributing independent circuit instances across the atom array, it is possible to increase throughput and reduce total execution time. Combined with the near-constant system power profile, this suggests that parallel execution can significantly improve energy efficiency by amortizing static costs across multiple simultaneous computations.

\vspace{2mm}

\noindent\textbf{Packing-Fidelity Trade-off.} Despite this opportunity, spatial parallelism is fundamentally constrained by interaction-induced noise. As atoms are placed closer together to increase packing density, the strength of residual Rydberg interactions between nominally independent circuit instances increases due to the $1/r_{ij}^6$ scaling. This leads to unintended entanglement, crosstalk, and degradation in computational fidelity.

Conversely, increasing the spacing between atoms suppresses these unwanted interactions and improves fidelity, but reduces the number of qubits that can be utilized concurrently, limiting throughput and increasing overall runtime and energy consumption. This establishes a fundamental trade-off between spatial resolution and computational accuracy in neutral-atom quantum computing systems.

\vspace{2mm}

\noindent\textbf{Motivation.} These observations highlight a missing systems-level abstraction: a unified framework that connects device-level interaction physics, spatial qubit allocation, and system-level metrics such as runtime, energy, and fidelity. Existing approaches typically treat fidelity, scheduling, and resource utilization independently, without translating their coupled dependence on physical layout and interaction strength into practical scheduling decisions.

\textit{This trade-off forms the central motivation for \sol{}, which translates the relationship between spatial packing, interaction-induced fidelity degradation, and system-level energy and runtime into a hardware-aware qubit packing policy for neutral atom quantum systems.}
\section{Power Consumption Model}
\label{sec:power}

The total power consumed by a neutral atom quantum computer arises from both static and dynamic components of the system. From a hardware and systems engineering perspective, these components correspond to distinct physical subsystems with different scaling behavior. The static component captures the constant optical and electronic infrastructure, including trapping lasers, optical tables, beam-delivery systems, and control electronics, which remain active regardless of the number of participating qubits~\cite{Pasqal_2023,boger2023dual}. The dynamic component captures the incremental power required to actively manipulate atoms during computation, including addressing beams, modulation hardware, and feedback/control bandwidth associated with gate execution and measurement~\cite{Boger_2025}.

We denote the total system power as:
\begin{equation}
P(n) = P_0 + P_\mathrm{dyn}(n),
\end{equation}
where $P_0$ is the static baseline power and $P_\mathrm{dyn}(n)$ represents the incremental power associated with operating $n$ active qubits. Because each qubit requires optical addressing and contributes to aggregate laser intensity, modulation bandwidth, and control complexity, $P_\mathrm{dyn}(n)$ scales approximately linearly with $n$ under fixed optical conditions.

However, an important architectural property of neutral atom systems is the extensive sharing of optical and control resources across qubits. Global laser drives, multiplexed beam steering, and shared control electronics significantly amortize per-qubit costs. As a result, platforms such as QuEra’s Aquila exhibit extremely favorable scaling behavior, where the marginal power required to control additional qubits is small relative to the static baseline. To capture this behavior at the systems level, we adopt the following generalized model:
\begin{equation}
P(n) = P_0 + \alpha n^\beta,
\label{eq:power_model}
\end{equation}
where $\alpha$ captures the effective per-qubit power contribution and $\beta$ models the scaling behavior of the control and optical subsystems. For current Rydberg implementations, $\beta \approx 1$ or slightly sub-linear due to multiplexed addressing and shared laser infrastructure.

\vspace{3mm}
\hrule
\vspace{1mm}

\noindent\textbf{Insight: Static-Power-Dominated Regime.} Neutral atom quantum computers operate in a regime fundamentally distinct from many classical and quantum architectures: system power is dominated by static overhead rather than dynamic per-qubit costs. For example, the QuEra Aquila system consumes approximately 7~kW at 256 qubits and is projected to require only around 10~kW at 10{,}000 qubits, despite the exponential increase in computational capacity~\cite{boger2023dual,Boger_2025}. 

Consistent with these observations, Eq.~\eqref{eq:power_model} captures a near-flat power scaling regime in which $P_0 \gg P_\mathrm{dyn}(n)$ across a wide range of system sizes. This regime has important implications for system-level execution: energy consumption becomes primarily a function of execution time rather than instantaneous qubit utilization. \textit{This static-power-dominated behavior motivates a shift in execution strategy.} In a serialized execution model, each circuit shot incurs the full baseline cost $P_0$ while utilizing only a fraction of available qubits, resulting in inefficient energy usage. In contrast, parallel execution allows multiple circuit instances to share the same static power cost within a single control cycle. By maximizing qubit utilization through spatial packing, the system can effectively amortize $P_0$ across many concurrent computations.

This leads to a key systems-level observation: in neutral-atom architectures, increasing parallelism can reduce both runtime and total energy, until interaction-induced fidelity degradation becomes dominant. In this sense, additional qubits are effectively ``free'' from an energy perspective, but not from a fidelity perspective.

\vspace{1mm}
\hrule
\vspace{3mm}

This observation forms the foundation for \sol{}'s fidelity-aware qubit packing. In Sec.~\ref{sec:bounds}, we use Eq.~\eqref{eq:power_model} to derive analytical bounds on runtime and energy under serial and parallel execution regimes. In Sec.~\ref{sec:design}, we incorporate spatial packing and interaction-induced noise to model realistic operating points that jointly balance energy, runtime, and fidelity.
\section{Bounds for Time and Energy}
\label{sec:bounds}

\begin{figure}
\centering
\includegraphics[width=0.99\linewidth]{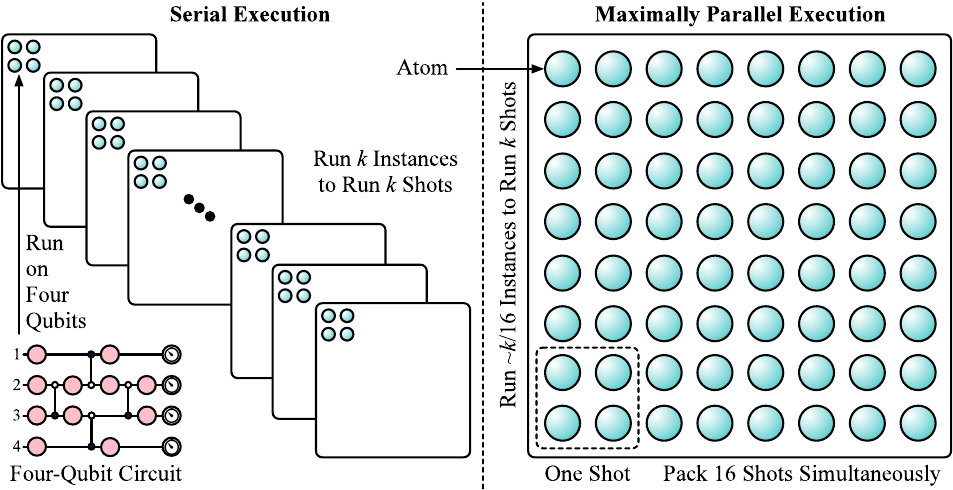}
\caption{Serial execution vs. maximally parallel execution.}
\label{fig:ideal}
\end{figure}

The total time and energy required to execute a quantum workload depend on how circuit instances are distributed across the available qubits. Let $n$ denote the number of qubits required by the circuit, $N$ the total number of available qubits on the device (typically $n << N$), $t$ the duration of a single shot of the algorithm, and $s$ the total number of shots to be executed. From a systems perspective, these parameters jointly determine how effectively the underlying hardware resources are utilized during execution.

\subsection{Serial Execution}

In the serial regime (Fig.~\ref{fig:ideal}), only one circuit shot is executed at a time using $n$ qubits. The total time required to complete all $s$ shots (one shot per executed ``instance'') is therefore:
\begin{equation}
\boxed{T_\mathrm{serial} = s \, t}
\label{eq:t_serial}
\end{equation}
The total energy consumption is obtained by integrating power over time, using the power model from Eq.~\eqref{eq:power_model}:
\begin{equation}
\boxed{E_\mathrm{serial} = T_\mathrm{serial} \, P(n)
                = s \, t \, \big(P_0 + \alpha n^{\beta}\big)}
\label{eq:e_serial}
\end{equation}

Equations~\eqref{eq:t_serial} and~\eqref{eq:e_serial} represent the upper bounds for total time and energy, as no parallelism is exploited. In this regime, each shot incurs the full system cost while utilizing only a fraction of the available qubits, resulting in low hardware utilization and inefficient amortization of the static power overhead. This is the prevailing method of execution.

\subsection{Maximally Parallel Execution}

If the full device capacity $N$ is available, multiple shots can be executed concurrently. The number of shots that can fit in one batch instance when packing maximally is given by:
\begin{equation}
k = \left\lfloor \frac{N}{n} \right\rfloor,
\end{equation}
and the number of batches required to complete all shots is $\lceil s/k \rceil$. In the idealized fully parallel case, the runtime therefore reduces to the following relationship:
\begin{equation}
\boxed{T_\mathrm{parallel} = \frac{s}{k} \, t \approx \frac{s \, n}{N} \, t}
\label{eq:t_parallel}
\end{equation}
As the power scales with the number of active qubits $N$ loaded into the atom array, the corresponding energy is:
\begin{equation}
\boxed{E_\mathrm{parallel} = T_\mathrm{parallel} \, P(N)
                   = \frac{s \, n}{N} \, t \, \big(P_0 + \alpha N^{\beta}\big)}
\label{eq:e_parallel}
\end{equation}

Equations~\eqref{eq:t_parallel} and~\eqref{eq:e_parallel} provide the lower bounds on total runtime and energy, assuming perfect packing and no interference between adjacent executions, even when densely packed (this is not a realistic setting). This regime corresponds to full hardware utilization, in which all available qubits contribute to computation in each execution cycle.

\subsection{Bound Interpretation}

Together, Equations~\eqref{eq:t_serial}-\eqref{eq:e_parallel} define a feasible operating envelope for any neutral atom quantum computer. These bounds characterize two extreme execution regimes: serialized execution, which maximizes runtime and energy while providing maximal isolation between circuit instances, and fully parallel execution, which minimizes runtime and energy under the assumption of perfect spatial packing and negligible interaction.

Real systems operate between these limits, where achievable packing efficiency is constrained by device-level interaction physics. As circuit instances are placed closer together, residual Rydberg interactions introduce crosstalk and degrade fidelity, preventing operation at the ideal parallel limit. Conversely, increasing spacing improves fidelity but reduces parallelism, moving the system toward the serial regime.

\textit{From a systems engineering perspective, these bounds expose a fundamental trade-off between hardware utilization and interaction-induced error.} While parallel execution improves throughput and amortizes static power, it also increases susceptibility to fidelity degradation. Therefore, the optimal operating point lies between these extremes and depends on the desired fidelity target. These effects are explicitly modeled next in Sec.~\ref{sec:design} through the proposed \sol{} framework, which translates fidelity constraints into packing decisions that jointly optimize runtime and energy while respecting interaction-induced crosstalk.

\section{\sol{}'s Formulation and Design}
\label{sec:design}

\begin{figure}
    \centering
    \includegraphics[width=0.99\linewidth]{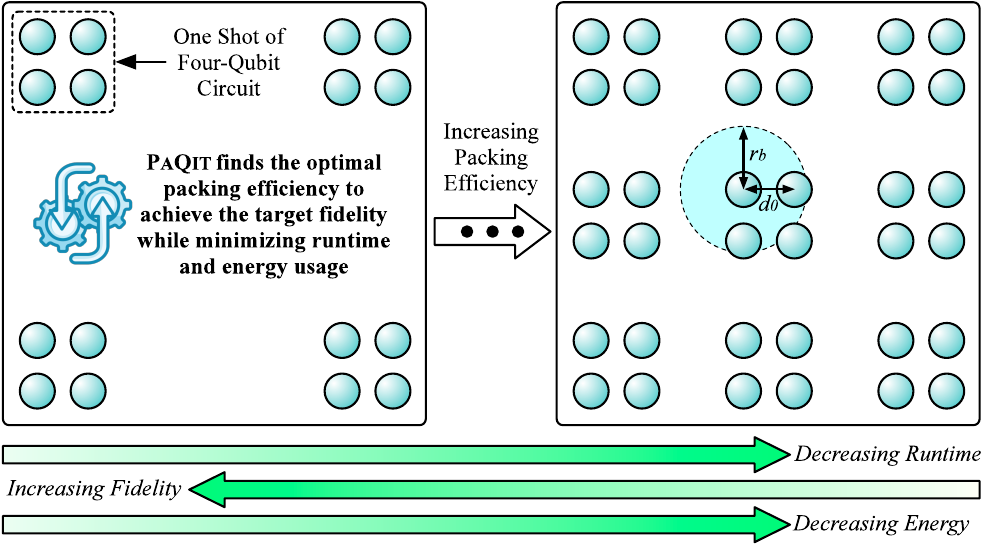}
    \caption{\sol{} determines the packing efficiency of the qubits on the neutral atom system to minimize the runtime and energy usage for a target output fidelity desired by the user.}
    \label{fig:packing}
\end{figure}

The \sol{} framework addresses the practical systems constraint that maximal spatial packing of quantum circuit instances on a neutral atom array is physically infeasible. When multiple shots are placed in close proximity within a single physical realization, residual Rydberg interactions introduce crosstalk, leading to degradation in gate fidelity. \sol{} explicitly models this interaction-driven trade-off and provides a tunable, hardware-aware packing policy that balances throughput, fidelity, and total energy (Fig.~\ref{fig:packing}).

\subsection{Motivation for \sol{}}

The Rydberg blockade phenomenon couples qubits through an interaction potential that decays with the sixth power of their separation. When two atoms $i$ and $j$ are separated by distance $r_{ij}$, the interaction energy is
\begin{equation}
V_{ij} = \frac{C_6}{r_{ij}^6}.
\end{equation}
If the excitation frequency shift $V_{ij}$ exceeds the Rabi drive amplitude $\Omega$, simultaneous excitation is suppressed, and the atoms are said to be ``blockaded.'' The corresponding blockade radius $r_b$ is given by the following expression~\cite{wurtz2306aquila}:
\begin{equation}
r_b = \left(\frac{C_6}{\sqrt{\Omega^2 + \Delta^2}}\right)^{1/6}.
\end{equation}
Atoms (i.e., qubits) separated by less than $r_b$, but not intending to interact with each other, cannot be reliably excited together, leading to correlated errors when shots are too densely packed. This establishes a direct link between spatial layout and error behavior, making packing a first-class control variable in neutral atom systems.

\subsection{Packing Efficiency and Fidelity}

Let $\eta \in (0,1]$ denote the \emph{packing efficiency}: the fraction of total qubits actively utilized in parallel execution. Perfect packing corresponds to $\eta=1$, while lower $\eta$ values represent spatial under-utilization introduced to maintain isolation.

We model the average inter-shot spacing $d$ as inversely proportional to $\eta^{1/2}$ in a 2-D square lattice:
\begin{equation}
d = d_0 \, \eta^{-1/2},
\end{equation}
where $d_0$ is the nominal lattice spacing under full occupancy in a 2D square-grid lattice.

The fidelity of a single shot, averaged over all qubits, decreases with increasing residual coupling to neighboring excitations. Using the $1/r^6$ interaction profile, the cumulative perturbation scales with $(r_b/d)^6$. Thus, the expected fidelity can be expressed as:
\begin{equation}
f(\eta) = f_\mathrm{max} \, \exp\!\left[-\kappa \!\left(\frac{r_b}{d_0}\right)^{6} \! \eta^{3}\right],
\label{eq:fidelity_eta}
\end{equation}
where $\kappa$ is a proportionality factor determined by the control pulse shape and detuning.

Eq.~\eqref{eq:fidelity_eta} captures the key systems-level behavior: increasing packing efficiency improves hardware utilization but induces exponential fidelity degradation due to interaction-driven crosstalk among the different circuit instances.

\subsection{Runtime-Fidelity Relationship}

Given a target fidelity $f^\ast$, the maximum allowable packing efficiency is obtained by inverting~\eqref{eq:fidelity_eta}:
\begin{equation}
\boxed{\eta(f^\ast) =
\left[
  -\frac{\ln(f^\ast / f_\mathrm{max})}
       {\kappa (r_b / d_0)^6}
\right]^{\!1/3}}
\label{eq:eta_f}
\end{equation}

Substituting $\eta(f^\ast)$ into the parallel runtime expression in~\eqref{eq:t_parallel}, the effective execution time under the target fidelity constraint becomes:
\begin{equation}
\boxed{T_\mathrm{eff}(f^\ast) =
\frac{s\,n}{\eta(f^\ast) N}\, t}
\label{eq:t_eff}
\end{equation}
Here, the denominator $\eta(f^\ast)N$ represents the number of qubits effectively available for concurrent execution.

Eq.~\eqref{eq:t_eff} reveals that runtime increases super-linearly as fidelity requirements tighten, since the usable fraction of the array decreases with $(\ln f^\ast)^{-1/3}$. When $\eta(f^\ast) = 1$, the runtime matches the maximally parallel regime in Eq.~\eqref{eq:t_parallel}. When $\eta(f^\ast) = \frac{n}{N}$, the runtime reduces to the serial regime in Eq.~\eqref{eq:t_serial}, providing a consistency check for the formulation.

\subsection{Energy-Fidelity Relationship}

Using the power model from Equation~\eqref{eq:power_model}, the total energy at a target fidelity is:
\begin{equation}
\boxed{E_\mathrm{eff}(f^\ast)
= T_\mathrm{eff}(f^\ast)
  \Big(P_0 + \alpha [\eta(f^\ast) N]^{\beta}\Big)}
\label{eq:e_eff}
\end{equation}
The first term in parentheses captures the constant overhead of trapping and control infrastructure, while the second term scales with the number of actively driven qubits. Because $T_\mathrm{eff}(f^\ast)$ increases as $\eta(f^\ast)$ decreases, and $P_\mathrm{dyn}$ simultaneously decreases, $E_\mathrm{eff}(f^\ast)$ exhibits a non-monotonic dependence on fidelity: dense packing reduces runtime but increases dynamic power, while sparse packing reduces dynamic power but increases runtime, causing static power to dominate.

\subsection{Optimization and Scheduling}

\sol{} identifies the operating point that minimizes total energy (equivalently, runtime in the static-power-dominated regime) for a given fidelity target, or maximizes fidelity under a fixed energy budget. Formally, the optimization problem is given by the following objective:
\begin{equation}
\min_{\eta \in (0,1]} E_\mathrm{eff}(f(\eta))
\quad
\text{s.t.} \quad
f(\eta) \ge f^*.
\label{eq:opt_problem}
\end{equation}

In practice, \sol{} can be implemented as a scheduler within the quantum control stack. Given a circuit’s qubit footprint $n$ and desired fidelity $f^\ast$, the scheduler computes $\eta(f^\ast)$ using Equation~\eqref{eq:eta_f} and allocates spatial regions on the neutral atom lattice such that concurrent shots maintain average spacing $d(\eta)$. This mapping directly translates device-level interaction physics into spatial layout constraints.

Note: \sol{} requires no hardware modifications and operates entirely at the software and control layout levels, making it compatible with existing neutral-atom platforms. The resulting configuration minimizes total energy while satisfying fidelity constraints and can be recomputed as workload requirements or hardware calibration parameters change. \textit{Through these formulations, \sol{} translates device-level Rydberg interaction physics into an actionable system-level control policy for jointly optimizing fidelity, runtime, and energy.}

\section{Implementation and Methodology}
\label{sec:method}

This section describes the parameterization and simulation setup used to evaluate \sol{} from a hardware-grounded, system-level perspective.

\subsection{Hardware Modeling}

We use a simulated 256-qubit QuEra \textit{Aquila} system for both experimental evaluation and parameter calibration, as it is the only publicly accessible analog neutral atom platform~\cite{wurtz2306aquila}. Aquila consists of a two-dimensional array of $N$ optical tweezers, each trapping a single $^{87}$Rb atom. Qubits are encoded in the ground $\ket{g}$, and Rydberg $\ket{r}$ states, coupled via a global Rabi drive $\Omega(t)$ and detuning $\Delta(t)$. Spatial selectivity is achieved through an acousto-optic deflector (AOD) and a spatial light modulator (SLM), enabling arbitrary atom placement and reconfigurable lattice geometries. This flexibility directly enables \sol{} to operate across a range of packing efficiencies under realistic hardware constraints.

The system parameters used in our model are derived from publicly reported Aquila characteristics~\cite{boger2023dual,Boger_2025}. These values are used to characterize relative system-level scheduling behavior rather than absolute device power consumption. Aquila operates at a static baseline power of $P_0 = 6.92~\mathrm{kW}$ and exhibits near-linear power scaling with active qubit count, modeled as
\begin{equation}
P(N) = P_0 + \alpha N^\beta, \text{where } \alpha = 0.000308~\mathrm{kW/qubit},~ \beta = 1.\notag
\end{equation}
This model reflects the shared optical and control infrastructure of neutral atom systems, where marginal per-qubit power remains low relative to the static baseline.

The lattice pitch $d_0 \approx 6~\mu$m and typical blockade radius $r_b \approx 10$–$15~\mu$m yield a normalized ratio $(r_b/d_0) \approx 1$ under nominal operation. Larger values correspond to stronger interaction regimes, reducing feasible packing density and directly constraining parallel execution.

To complement the analog setting, we also evaluate \sol{} using simulations of the recently introduced QuEra \textit{Gemini} digital neutral atom system~\cite{Sales_Rodriguez_2025}. Gemini comprises $260$ physical $^{87}$Rb qubits arranged in a dynamically reconfigurable array, with atoms shuttled between storage and entangling zones. This architecture provides effective all-to-all connectivity and supports parallel two-qubit CZ gates. Reported native fidelities exceed $99.7\%$ for single-qubit gates and $99.2\%$ for parallel two-qubit gates, enabling controlled study of circuit-structure-dependent noise effects independent of spatial packing.

\subsection{Simulation Framework}

We implement a discrete-event simulator in Python~3.11 that integrates the analytical models derived in Sections~\ref{sec:power}–\ref{sec:design}. The simulator captures the mapping from fidelity targets to packing configurations, and from packing to runtime and energy, enabling end-to-end evaluation of \sol{}.

Each run simulates $s=1000$ circuit shots, each using $n=16$ qubits on a system with $N=256$ total qubits and executing for $t=16~\mu$s. These parameters are configurable and can be adjusted to reflect different workloads, hardware scales, and target fidelity requirements. For each configuration, the simulator computes the following system-level metrics as a function of target fidelity $(f^\ast)$:

\begin{itemize}
    \item \textbf{Packing efficiency} $\eta(f^\ast)$ from Eq.~\eqref{eq:eta_f},

    \vspace{1mm}
    
    \item \textbf{Runtime} $T_\mathrm{eff}(f^\ast)$ from Eq.~\eqref{eq:t_eff},

    \vspace{1mm}
    
    \item \textbf{Energy} $E_\mathrm{eff}(f^\ast)$ from Eq.~\eqref{eq:e_eff}.
\end{itemize}

The physical scaling constants are set to $\kappa=1.0$ and $f_\mathrm{max}=1.0$, representing normalized interaction sensitivity under typical control pulse configurations. While simplified, these normalized parameters allow us to isolate and study the structural dependence of system behavior on packing and interaction strength independent of hardware-specific calibration.

\subsection{Parameter Sweeps}

To characterize the operating regimes of \sol{}, we perform parameter sweeps over fidelity and interaction targets:
\begin{itemize}
    \item Target fidelity $f^\ast$ from $0.8$ to $1.0$,
    \item Normalized blockade ratio $r_b/d_0 \in \{0.5, 1.0, 2.0\}$.
\end{itemize}

All other parameters remain fixed. These sweeps generate three primary curves, packing efficiency, runtime, and energy as functions of fidelity, presented in Sec.~\ref{sec:evaluation}.

Varying $(r_b/d_0)$ allows us to emulate different hardware operating regimes, from weakly interacting systems that permit dense packing to strongly interacting regimes where spatial isolation is required. This directly exposes how interaction physics constrains achievable parallelism and drives system-level trade-offs between fidelity, runtime, and energy.

\subsection{Implementation and Reproducibility}

All experiments were conducted on a standard Apple workstation equipped with an M1 chip, 8 CPU cores, and 8 GB of RAM. The full parameter sweep completes in under two minutes, enabling rapid exploration of operating regimes. All simulation scripts use \texttt{NumPy} and \texttt{Matplotlib} for numerical analysis and visualization. Source code and parameter configurations will be released publicly upon publication to ensure full reproducibility of \sol{}’s evaluation.

\vspace{2mm}

\noindent\textbf{Noisy Analog Implementation.} To model interaction-driven effects in neutral atom systems, we use the Bloqade analog Python emulator to simulate a square array of atoms driven by a global Rydberg pulse. The atoms are arranged on a square lattice with spacing $a$ (in $\mu\mathrm{m}$), yielding a total of $N = \text{side}^2$ qubits. For each configuration $(\text{side}, a)$, we apply a uniform Rydberg drive with a piecewise-linear ramp–hold–ramp profile. The Rabi frequency $\Omega(t)$ starts at zero, increases to a maximum value $\Omega_\text{max} = 5~\mathrm{rad}/\mu\mathrm{s}$, remains approximately constant, and then ramps back to zero. The segment durations are chosen such that the total pulse area $\int \Omega(t)\,dt$ matches that of an ideal flat $\pi/2$ pulse, ensuring that a single non-interacting atom yields an excited-state probability of $p_e \approx 0.5$.

We simulate time evolution under this control pulse, sample measurement outcomes, and extract the per-site Rydberg excitation densities $\{\rho_i\}_{i=1}^N$. We define fidelity as:
\[
F = 1 - \frac{1}{p_\text{ideal}} \cdot \frac{1}{N} \sum_{i=1}^N \bigl|\rho_i - p_\text{ideal}\bigr|,
\]
with $p_\text{ideal} = 0.5$. Under this definition, $F = 1$ corresponds to independent, non-interacting qubits, and deviations capture interaction-induced distortion due to residual Rydberg coupling. This metric directly reflects the impact of spatial packing on the behavior of analog systems. We intentionally use this state-preparation experiment to isolate interaction-induced crosstalk from algorithm-specific compilation effects, allowing the underlying packing behavior modeled by \sol{} to be evaluated directly.

\vspace{2mm}

\noindent\textbf{Real Aquila Hardware Execution.} To validate that the interaction-driven packing trends assumed by \sol{} persist on real neutral-atom hardware,, we execute the same analog state-preparation experiment on QuEra Aquila through AWS Braket. We submit analog Hamiltonian simulation tasks for square arrays with side length $m \in \{1,2,3,4\}$, corresponding to $N=m^2$ atoms, and sweep nearest-neighbor lattice spacing $a \in \{4,5,6,7,8,9,10,11\}\,\mu\mathrm{m}$. Each configuration is executed for $1000$ shots using the same global ramp--hold--ramp Rydberg drive as the simulator, with $\Omega_\mathrm{max}=5\,\mathrm{rad}/\mu\mathrm{s}$ and pulse area chosen to approximate a single-atom $\pi/2$ excitation.

Aquila reports pre- and post-sequence occupation measurements for each shot. We count site $i$ as successfully loaded in a shot when its pre-sequence value is occupied, and count it as Rydberg-excited when it was successfully loaded before the pulse and measured as absent afterward. Let $L_i$ denote the number of successfully loaded shots at site $i$, and $R_i$ denote the number of those shots ending in the Rydberg state. The measured per-site Rydberg density is therefore $\rho_i = R_i/L_i$. We then compute the same fidelity proxy used in the analog simulations, with $p_\text{ideal}=0.5$.

This is not full quantum state or process fidelity; rather, it is a state-preparation fidelity proxy that measures how closely the real hardware preserves the independent single-atom response expected under the global pulse. Deviations from $p_\text{ideal}$ capture both interaction-induced suppression from dense spatial packing and real-device noise sources such as loading, control, and measurement imperfections.

\vspace{2mm}

\noindent\textbf{Noisy Digital Implementation.} To isolate the effect of circuit structure under fixed hardware noise, we perform a complementary study using simulations of QuEra’s digital Gemini architecture. Using the \texttt{GeminiOneZoneNoiseModel} from Bloqade, we simulate circuits on $N \in \{2,\dots,9\}$ line qubits and compute the Hilbert–Schmidt fidelity $F = \mathrm{Tr}(\rho_{\mathrm{ideal}}\rho_{\mathrm{noisy}})$ between the ideal and noisy output states after compilation to Gemini’s native CZ+PhXZ gate set. To span a range of workload characteristics under a consistent noise model, we evaluate four circuit families: (1) a GHZ chain, constructed via a nearest-neighbor CNOT cascade, (2) a Bell-pair circuit consisting of disjoint nearest-neighbor entangled pairs, (3) a brickwork CZ circuit with alternating layers of nearest-neighbor entangling gates applied to an initial $|+\rangle^{\otimes N}$ state, and (4) a fully connected CZ circuit that applies entangling gates between all qubit pairs, resulting in $O(N^2)$ interactions. These circuit families systematically vary entangling density while holding the noise model fixed, allowing us to isolate how circuit structure alone influences fidelity degradation. Together with the analog simulations, this provides a unified view of how spatial packing and circuit-level complexity jointly affect performance in neutral-atom systems, illustrating how \sol{} can support a range of workload characteristics beyond the state-preparation experiments.
\section{Evaluation and Discussion}
\label{sec:evaluation}

\begin{figure}[t]
    \centering
    \includegraphics[width=0.99\linewidth]{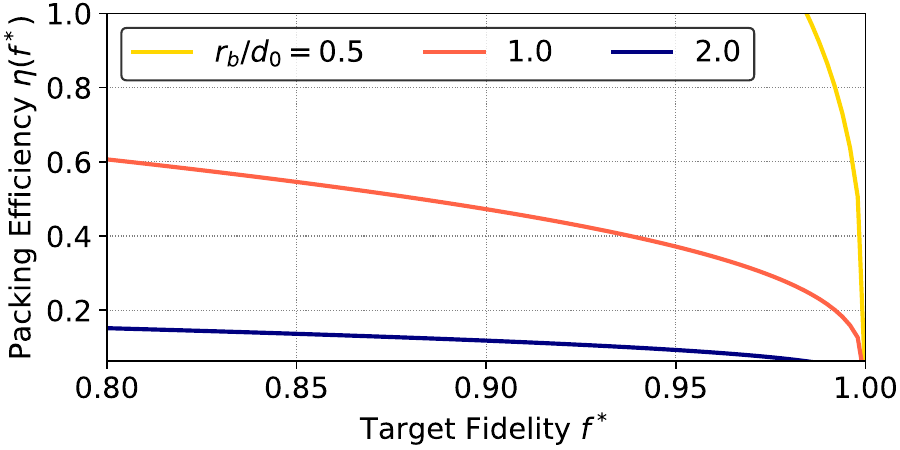}
    \caption{The packing efficiency decreases with increasing fidelity, with strong dependency on the $\frac{r_b}{d_0}$ relationship.}
    \label{fig:fid_pack}
\end{figure}

\begin{figure}
    \centering
    \includegraphics[width=0.99\linewidth]{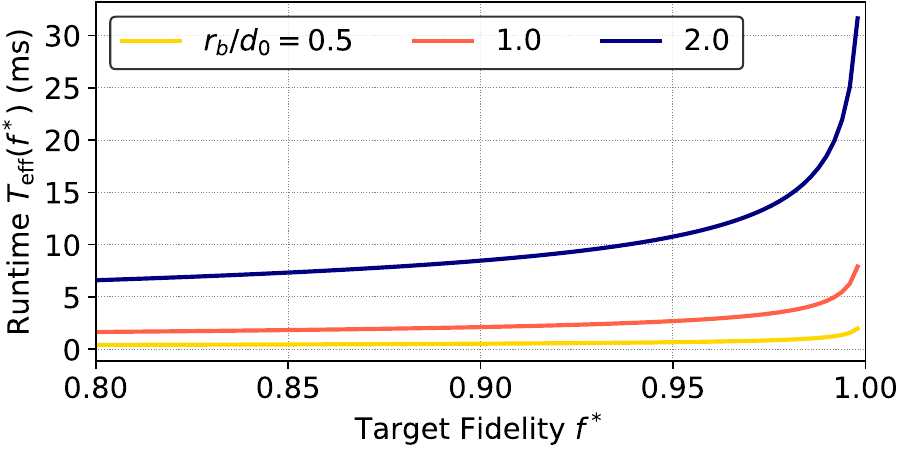}
    \caption{The total runtime of a circuit increases with higher target fidelity, with strong dependency on the $\frac{r_b}{d_0}$ relationship.}
    \label{fig:fid_time}
\end{figure}

\subsection{Trends Revealed by \sol{}'s Model}

\begin{figure}
    \centering
    \includegraphics[width=0.99\linewidth]{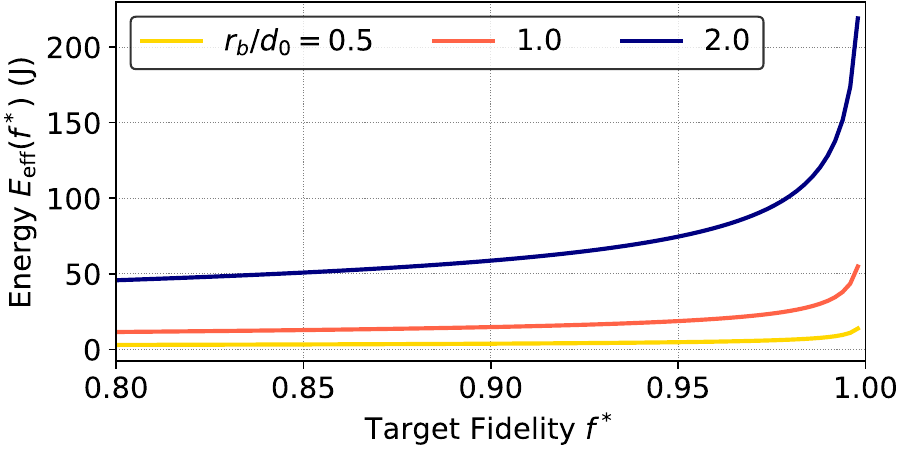}
    \caption{The energy consumption also increases with increasing fidelity, with strong dependency on the $\frac{r_b}{d_0}$ relationship.}
    \label{fig:fid_energy}
\end{figure}

Figures~\ref{fig:fid_pack}-\ref{fig:fid_energy} collectively illustrate the relationships among metrics across hardware operating regimes for neutral-atom architectures in an idealized setting. These results provide a system-level characterization of how spatial utilization interacts with interaction-induced noise and static-power-dominated execution. The analytical trends in Figs.~\ref{fig:fid_pack}-\ref{fig:fid_energy} illustrate the operating regimes predicted by \sol{}'s formulation. The experimental results in Figs.~\ref{fig:aquilaruns}-\ref{fig:geminiruns} then validate the underlying physical assumptions of the framework, namely that denser spatial packing increases interaction-induced crosstalk, while larger spacing preserves fidelity at the expense of parallelism.

\vspace{2mm}

\noindent\textbf{Packing Efficiency vs. Fidelity.} Fig.~\ref{fig:fid_pack} shows that packing efficiency $\eta(f^\ast)$ drops steeply as the target fidelity increases. The rate of decline is highly sensitive to the normalized blockade ratio $(r_b/d_0)$, which captures the relative strength and spatial extent of Rydberg interactions. A larger ratio indicates stronger, longer-range interactions, effectively creating a broader exclusion region around each excited atom. As $(r_b/d_0)$ increases, qubits must be placed farther apart to suppress interaction-induced crosstalk, directly reducing achievable spatial utilization.

In contrast, smaller ratios correspond to weaker interactions, allowing denser packing with limited fidelity degradation. This highlights a key hardware-dependent parameter that governs the system's achievable operating regime. This ratio captures a fundamental trade-off between circuit density and robustness. Algorithms dominated by local gates or shallow variational layers can safely operate at lower $d_0$, leveraging dense packing to improve throughput. Conversely, circuits with frequent entangling operations or high interaction depth require larger $d_0$ or reduced drive strength to avoid cross-excitation.

Decreasing the Rabi frequency $\Omega$ enlarges the blockade radius $r_b = (C_6 / \hbar \Omega)^{1/6}$, further tightening packing constraints. Thus, \sol{} exposes a direct coupling between control parameters (e.g., $\Omega$, $\Delta$) and spatial scheduling decisions, making packing efficiency a tunable system-level knob.

\vspace{2mm}

\noindent\textbf{Runtime vs. Fidelity.} Fig.~\ref{fig:fid_time} shows that total runtime $T_\mathrm{eff}(f^\ast)$ increases as fidelity targets become more stringent. As packing efficiency decreases, fewer circuit instances can be executed concurrently, forcing partial or full serialization of shots.

This transition from parallel to serial execution is governed by the same interaction constraints captured by $(r_b/d_0)$. For larger ratios, even modest increases in fidelity requirements lead to sharp reductions in parallelism, resulting in significant runtime inflation. This behavior confirms that throughput in neutral-atom systems is fundamentally limited by interaction-induced constraints rather than by raw qubit count.

From a systems perspective, this implies that increasing hardware scale alone is insufficient to guarantee improved performance; effective utilization depends critically on maintaining sufficient spacing to control interaction errors.

\vspace{2mm}

\noindent\textbf{Energy vs. Fidelity.} Fig.~\ref{fig:fid_energy} presents the corresponding total energy trend. Because neutral atom systems operate in a static-power-dominated regime, total energy is strongly correlated with execution time. At lower-fidelity targets, dense packing enables high parallelism, reducing both runtime and dynamic power consumption. As fidelity requirements increase, packing efficiency decreases, leading to longer runtimes and under-utilization of the hardware. In this regime, static power $P_0$ dominates, and energy consumption increases sharply.

The resulting convex profile of $E_\mathrm{eff}(f^\ast)$ reveals the existence of an optimal operating point that balances parallelism and fidelity. This point corresponds to the regime where marginal gains in fidelity begin to incur disproportionate increases in runtime and energy. \sol{} explicitly identifies this operating point, enabling principled selection of execution configurations based on system-level objectives.

\subsection{Noisy Executions on Aquila and Gemini}

\begin{figure}[t]
    \centering
    \includegraphics[width=0.99\linewidth]{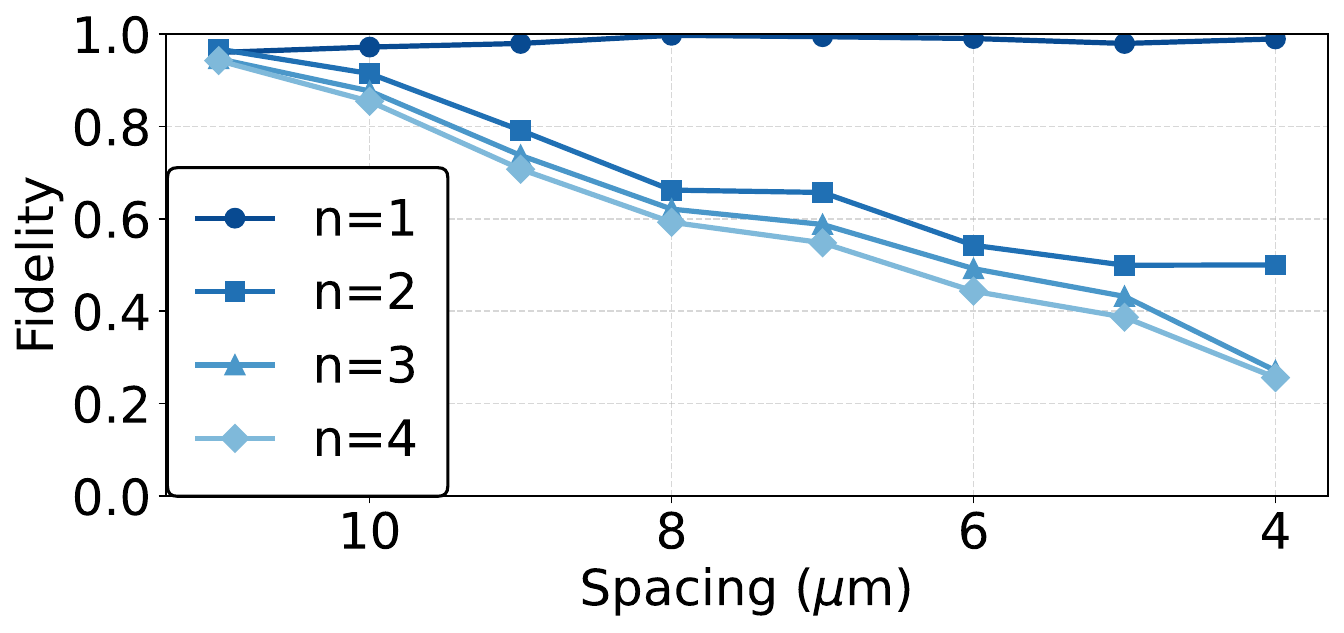}
    \caption{Simulation on an Aquila array showing state-preparation fidelity versus atom spacing for four array sizes.}
    \label{fig:aquilaruns}
\end{figure}

\noindent\textbf{Qubit Count and Spacing vs. Fidelity.} Fig.~\ref{fig:aquilaruns} shows that fidelity remains near unity when atoms are well separated or when only a single atom is loaded, but systematically degrades as lattice spacing is reduced and as more atoms are packed into the array. For Aquila, global drive of up to \(\Omega_{\max}=1.58\times 10^7\,\mathrm{rad/s}\) and detuning \(|\Delta_{\max}|=1.25\times 10^8\,\mathrm{rad/s}\), with minimum site spacing \(d_0 \approx 4~\mu\mathrm{m}\), which together imply a characteristic blockade distance \(r_b \sim 6\text{--}8~\mu\mathrm{m}\), i.e., \(r_b/d_0 \sim 1.5\text{--}2\) in the densely packed regime probed by these experiments~\cite{wurtz2306aquila}.  As spacing is reduced, the $n=2$-4 curves drop by approximately a factor of two to three, demonstrating that interaction-induced errors accumulate rapidly even at small system sizes. This empirically validates the central assumption underlying \sol{}'s analytical formulation: aggressive spatial packing directly trades off fidelity against utilization due to residual Rydberg coupling.

Importantly, these results confirm that the assumptions underlying the analytical model capture the qualitative behavior observed in realistic analog hardware, reinforcing the validity of using $(r_b/d_0)$ as a key control parameter within \sol{}.

\vspace{2mm}

\begin{figure}[t]
    \centering
    \includegraphics[width=0.99\linewidth]{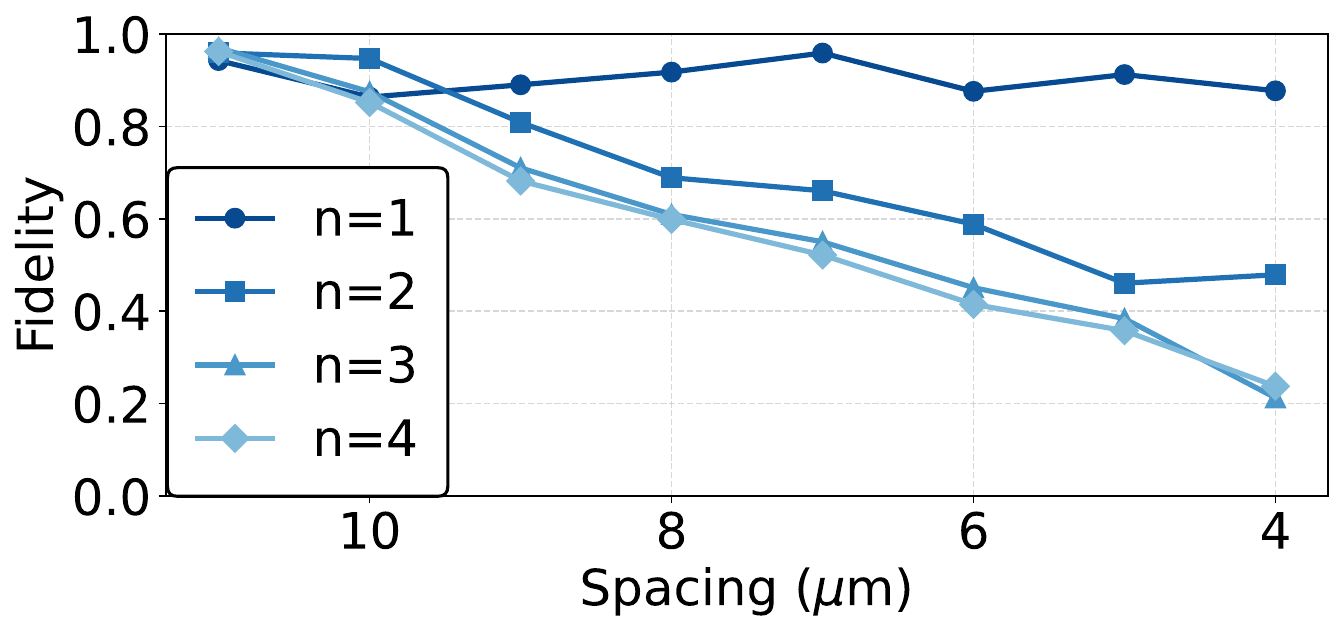}
    \caption{Real Aquila hardware executions showing state-preparation fidelity versus atom spacing for four array sizes.}
    \label{fig:aquilarealruns}
\end{figure}

\noindent\textbf{Real Aquila Hardware Validation.} Fig.~\ref{fig:aquilarealruns} shows the corresponding results from real Aquila hardware executions. The measured state-preparation fidelity proxy follows the same qualitative trend as the simulated Aquila data: when atoms are separated by $10$--$11\,\mu$m, all array sizes exhibit similar behavior, indicating little measurable interference between atoms, regardless of the number of atoms overall in the grid. As spacing is reduced below this regime, the fidelity proxy decreases approximately monotonically, consistent with stronger residual Rydberg interactions under denser packing.

The real hardware data also exhibits additional incoherent variation relative to simulation, as expected from loading, control, and measurement imperfections. This is most visible in the $n=1$ reference curve, which fluctuates around the $0.9$ fidelity level despite having no neighboring atoms with which to interact. Thus, deviations in the single-atom case primarily reflect hardware noise rather than packing-induced crosstalk, and give a baseline expectation for crosstalk-free fidelity of qubits in Aquila.

Despite this added noise, the higher-density behavior remains consistent with the simulation. The $2\times2$ array shows a clear fidelity drop as spacing decreases, including a plateau between $5$ and $4\,\mu$m that also appears in the simulated data. More notably, while fidelity drops substantially when moving from the $2\times2$ to the $3\times3$ setting, the $3\times3$ and $4\times4$ curves remain close across most spacings. \textbf{This suggests that, assuming spacing between atoms is held fixed and that the atom grid is large enough to contain the spaced-atoms, adding additional atoms to the array does not necessarily introduce a proportional fidelity penalty}. For \sol{}, this reinforces the view that spatial separation, rather than raw atom count alone, is the dominant scheduling variable: if the hardware provides enough area to preserve spacing, more qubits can be packed into the computation with little additional loss in this fidelity proxy.

\vspace{2mm}

\begin{figure}[t]
    \centering
    \includegraphics[width=0.99\linewidth]{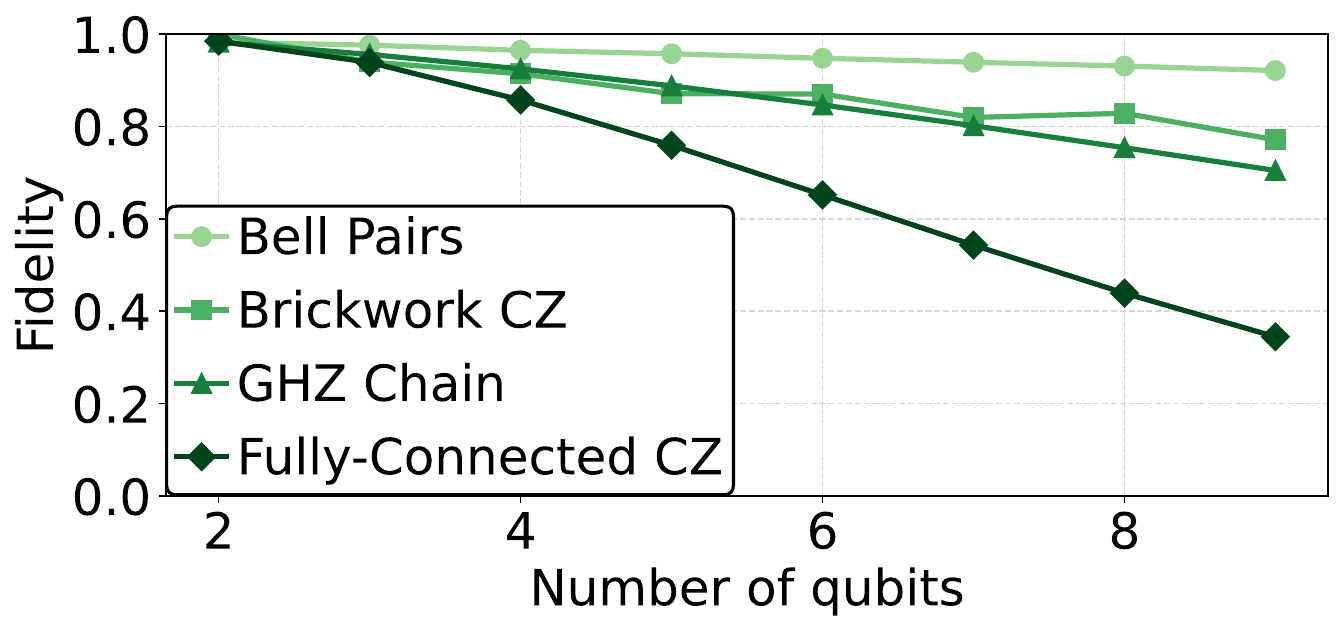}
    \caption{Gemini one-zone simulation of state-preparation fidelity versus qubit count for four circuit families.}
    \label{fig:geminiruns}
\end{figure}

\noindent\textbf{Qubit Count and Circuit Family vs. Fidelity.} Fig.~\ref{fig:geminiruns} shows that, under the Gemini noise model, fidelity degrades systematically with both qubit count and entangling density of the workload. The Bell-pair circuit, which contains the fewest two-qubit gates, maintains the highest fidelity across all sizes and remains close to 0.9 even at nine qubits. In contrast, the fully connected CZ circuit, which applies $O(N^2)$ entangling operations, exhibits the steepest decay, dropping to roughly one third fidelity at the largest $N$.

These results demonstrate that circuit structure alone can induce significant variation in fidelity, even in the absence of explicit spatial packing constraints. In particular, entangling density acts as an implicit amplifier of noise, compounding the effects of hardware-level imperfections.

Overall, the ordering of these circuit families under a realistic digital noise model provides a complementary baseline to the analog results, and is consistent with the trends predicted by \sol{}. Together, these experiments illustrate that the proposed scheduling policy naturally extends across workloads with different circuit structures and fidelity requirements, while highlighting that both spatial layout and circuit structure must be considered jointly when optimizing execution on neutral-atom platforms.

\vspace{2mm}

\noindent\textbf{Takeaways for Packing-Aware Scheduling.}
Our results above on Aquila suggest that packing decisions should be guided primarily by spatial separation rather than by qubit count alone. In both simulation and hardware, the fidelity proxy degrades sharply as the nearest-neighbor spacing is reduced, but the difference between the $9$ and $16$ atoms is comparatively small when the spacing is held fixed. This indicates that, beyond a modest array size, adding more atoms does not necessarily introduce a proportional fidelity penalty. Instead, the dominant factor is whether the physical layout maintains sufficient distance between atoms driven simultaneously.

This observation is important for \sol{} because it supports a scheduler-level abstraction based on spacing-constrained regions rather than conservative limits on total active qubits. A naive policy might reduce parallelism simply because more atoms are active in the same execution window.

Our results suggest a more favorable regime in which, if the device has sufficient physical area to maintain spacing above the interaction-sensitive threshold, additional circuit instances can be packed with minimal additional fidelity loss. In static-power-dominated neutral-atom systems, this creates an opportunity to increase throughput and reduce energy consumption while preserving accuracy.

More broadly, these results motivate treating neutral-atom arrays as spatial resources rather than only as collections of qubits. The relevant scheduling question is not only how many atoms are active, but also how they are geometrically arranged relative to the blockade radius and control pulse parameters. This reinforces our central design principle: energy-efficient execution should maximize parallelism subject to hardware-grounded spacing constraints, using application-specific fidelity targets to determine when dense packing is beneficial and when isolation is required.

\section{Related Work}

\label{sec:related}

Prior research on sustainable quantum computing has primarily focused on lifecycle carbon footprint, data center integration, resource optimization, and the potential long-term benefits of quantum acceleration~\cite{arora2024sustainable,auffeves2022quantum,deng2024power}. Several studies have argued that quantum computing may offer intrinsic energy advantages compared to classical high-performance computing under specific workload and scaling conditions~\cite{boger2023dual,jaschke2023quantum,Boger_2025}. However, these works largely operate at a system-agnostic level and do not provide hardware-grounded, analytically derived models that connect energy consumption to execution dynamics in programmable neutral-atom quantum computing platforms.

On the systems and programmability side, neutral-atom platforms have rapidly advanced through improvements in control, entanglement transport, and scalable architecture design, supported by both experimental and compiler-level innovations~\cite{bluvstein2022quantum,balewski2024engineering}. Recent compiler and architecture efforts have explored constraint-aware scheduling, routing, and parallelization for Rydberg hardware, demonstrating the importance of spatial layout and field-programmable configurations in maximizing throughput~\cite{ludmir2024parallax,wang2024atomique,tan2025compilation}. These works highlight the role of spatial resource allocation but primarily focus on performance metrics such as latency and circuit fidelity.

Related work has also examined hardware–software co-design and benchmarking using simulations that capture realistic neutral-atom dynamics~\cite{wang2024q,ludmir2024modeling,tan2022qubit}. These efforts provide valuable insights into device behavior and execution efficiency, but do not explicitly model energy as a first-class system metric or analyze its interaction with runtime and fidelity. Further, recent Rydberg-focused execution frameworks have demonstrated that analog, parallel workloads can be reconfigured or repurposed for advanced quantum machine learning and neural processing tasks, suggesting opportunities for spatially efficient execution patterns that directly interact with underlying hardware physics~\cite{dibrita2024recon,dibrita2025resq}. 

\textit{However, none of the works discussed above translate these hardware-level trade-offs into a fidelity-aware qubit packing policy that jointly optimizes energy, runtime, and fidelity under hardware constraints. This gap motivates the analytical framework and scheduling policy introduced in \sol{}.}
\section{Conclusion}
\label{sec:conclusion}

This work introduced \sol{}, a hardware-aware qubit-packing framework for Rydberg-atom quantum computers that translates fidelity constraints into scheduling decisions, jointly optimizing runtime and energy consumption. By modeling the \(1/r_{ij}^6\) interaction potential and deriving analytical bounds for power, time, and energy, \sol{} identifies operating points that balance spatial utilization with interaction-induced error within the system. Our results show that neutral-atom systems operate in a static-power-dominated regime, where parallel execution can significantly reduce both runtime and total energy until fidelity degradation limits performance. Because the framework is driven by user-specified fidelity targets, it can naturally adapt to different applications and workload requirements without requiring hardware modifications. \textit{\sol{} establishes a foundation for energy-aware, hardware-grounded quantum system design, enabling principled co-optimization of fidelity, runtime, and energy in neutral-atom architectures.}

\section*{Acknowledgement} 

We thank the anonymous reviewers for their helpful comments, which helped improve this work. This work was supported by the U.S. Department of Energy, Office of Science, National Quantum Information Science Research Centers, Quantum Science Center. This work was supported by Rice University, the Rice University George R. Brown School of Engineering and Computing, and the Rice University Department of Computer Science.

This work was also supported by the Ken Kennedy Institute and the Rice Quantum Initiative, which is part of the Smalley-Curl Institute. We also acknowledge the support of AWS Braket Cloud and QuEra Aquila for this work.

\balance

\bibliographystyle{IEEEtran}
\bibliography{main}

\end{document}